\documentclass[11pt]{article}
\usepackage{times}
\usepackage{url}
\usepackage{latexsym}
\usepackage{color}
\usepackage{graphicx}
\usepackage{lipsum}
\usepackage{natbib}
\usepackage[margin=1.0in]{geometry}

\title{You Are What You Eat... Listen to, Watch, and Read}

\author{Mason Bretan \\
  Georgia Institute of Technology\\
  {\tt masonbretan@gmail.com}
}

\date{}

\begin{document}
\maketitle
\begin{abstract}
This article describes a data driven method for deriving the relationship between personality and media preferences. A qunatifiable representation of such a relationship can be leveraged for use in recommendation systems and ameliorate the ``cold start" problem. Here, the data is comprised of an original collection of 1,316 Okcupid dating profiles. Of these profiles, 800 are labeled with one of 16 possible Myers-Briggs Type Indicators (MBTI). A personality specific topic model describing a person's favorite books, movies, shows, music, and food was generated using latent Dirichlet allocation (LDA). There were several significant findings, for example, intuitive thinking types preferred sci-fi/fantasy entertainment, extraversion correlated positively with upbeat dance music, and jazz, folk, and international cuisine correlated positively with those characterized by openness to experience. Many other correlations confirmed previous findings describing the relationship among personality, writing style, and personal preferences. (For complete word/personality type assocations see the Appendix).
\end{abstract}


\section{Introduction}
Recommendation systems have become increasingly popular over the last several years. These systems are catered towards a multitude of mediums including music (Pandora, Last.FM), movies (Netflix), literature (Amazon), food and restaurants (Yelp, Urbanspoon), and people (Okcupid). Many of these systems rely on user history within the specific domain combined with collaborative filtering techniques to recommend new options. For example, Netflix uses a user's previous views to create suggestions. Pandora uses a thumbs up and down rating system from users to inform their playlists. Other systems such as Amazon suggest content based off of purchases from similar buyers.

These methods have proven to be relatively successful, however, we ponder if additional information about users not directly related to the content can be useful for predicting his or her preferences. Additionally, it is difficult for current systems to make recommendations to first time users when zero previous data is available to inform the system of what a person might buy, watch, or listen to. Netflix may recommend movies and television shows using what is currently popular, which seems like a logical approach, but if a user is outside of the norm she may be turned off to the suggestions and first impressions can be tantamount to an application's success. Ideally the system should leverage some prior knowledge about a user before making suggestions to ameliorate this ``cold start" problem. In this work we examine traits that are descriptive of a person and correlate with a person's preferences. Specifically, we examine the relationship between an individual's personality and her preferences in music, movies, TV shows, food, and literature.

\section{Objectives}
There is a plethora of work examining the relationship between personality and media preferences. Personality traits such as psychoticism and neuroticism have been linked to preferences for horror/violent films and documentaries respectively \citep{weaver1991exploring}. In a more recent study Rentfrow found additional correlations between media preferences, personality, and geographical region \citep{rentfrow2011listening}. Though the question of whether the type of media an individual consumes influences his or her personality or if he or she is predisposed to liking specific entertainment as a result of personality is still being argued, the evidence that there exists a relationship between personality and preference is undeniable. This relationship is even leveraged in interpersonal communication between two strangers by asking questions about musical tastes to establish an impression of one another's personalities \citep{rentfrow2006message}.

The majority of the work examining the relationship between personality and preferences has been carried out using controlled user studies with surveys and questinonaires. In this work we attempt to reaffirm the previous findings by examining data collected from the online dating site {\it Okcupid}. In order to find any relationships between personality and aesthetic preferences we must have a method for identifying an individual's personality and that person's preferences. Therefore, there are four main objectives in this work:

\begin{description}
  \item[First] Construct a dataset allowing us to examine both personality and personal preferences
  \item[Second] Identify bag-of-word features indicative of different personality traits
  \item[Third] Identify personality specific bag-of-words describing aesthetic preferences
  \item[Fourth] Identify general topics acrross all personality traits and correlate them with individual personality traits
\end{description}

In the folowing sections we discuss the methods and results for each of these four goals.

\section{Okcupid Dataset}
Online dating profiles provide a good source of data for our purposes. Users are encouraged to give full and rich descriptions about themselves in order to paint a picture of the type of person he or she is. A typical Okcupid profile consists of multiple sections which help to define the user. These sections include:

\begin{itemize}
  \item self summary
  \item what I'm doing with my life
  \item I'm really good at
  \item the first things people notice about me
  \item {\color[rgb]{0,0,1} favorite books, movies, shows, music, and food}
  \item six things i could never do without
  \item on a typical Friday night I am
  \item I'm looking for
  \item you should message me if
\end{itemize}

The objective is to find correlations between the content provided in the {\it favorite books, movies, shows, music, and food} section and the information provided in the other sections of a user’s profile. The dataset consists of 516 female profiles. The profiles were selected at random, however, there were two constraints. The profile had to be within the USA (this does not mean the user is a US citizen, but rather increases the likelihood that the profile is in English). Secondly, the profile had to have completed the ``favorite books, movies, shows, music, and food” section (completing the other sections was not required).

In addition to the 516 random profiles another personality-labeled dataset was also constructed. Many users include their Myers-Briggs Type Indicator (MBTI) in their profiles. An MBTI label consists of four dichotomies representing a person's attitude (\textbf{E}xtraversion, \textbf{I}ntroversion), perceiving function (\textbf{S}ensing, i\textbf{N}uition), judging function (\textbf{T}hinking, \textbf{F}eeling), and lifestyle (\textbf{J}udging, \textbf{P}erception). Using Okcupid's keyword search function 800 MBTI-labeled profiles were collected with 50 for each of the 16 distinct types (ENTJ, ESTJ, INFP, etc.).

McCrae and Costa showed that the MBTI scales correlate with the ``Big Five" personality traits with attitude, perceiving, judging, and lifestyle correlating with extraversion, openness, agreeableness, and conscientiousness respectively \citep{mccrae1989reinterpreting}. There was no correlation with the fifth big five trait of neuroticism, however.

\section{Finding Personality Descriptors}
\subsection{Methods}
In the first experiment the goal is to find words in Okcupid profiles that are good indicators of each of the MBTI traits. To do this the MBTI-labeled dataset was used. First the dataset was split into two groups according to the attitude trait (extraversion vs introversion). Topic modeling using trait specific LDA was performed so that word lists unique to the two groups could be generated. This whole process of splitting the dataset into two groups and performing trait specific LDA was repeated three more types for the remaining MBTI dichotomies.

\subsection{Results and Discussion}
The top 10 words from each trait topic (not including the MBTI keywords) are shown in table 1. More complete lists of words and their respective weights are shown in Figures 1-4 in the appendix.

\begin{table*}[t]
\begin{center}
\begin{tabular}{|c|c|c|c|c|c|c|c|}
\hline \bf E & \bf I & \bf S & \bf N & \bf T & \bf F & \bf J & \bf P \\ \hline
beer & intelligence & bacon & yourself & intelligence & singing& intelligence & drinking\\
wine & quiet & intelligence & intellectual &his &she & his & shit\\
funny & her & video & our &unless& strips & bacon & creative\\
outgoing & such & guy & deep &romantic& bacon & graduate & name\\
sports & shy & name & ideas & smart&laughing & bacon & ass\\
moved & bacon  & sleep & walking & women&listening & dinner & used\\
drink & she & strips & change & emotional&cats& guy & notice\\
phone & alone & netflix & universe&several&figure & yes & couple\\
active & strips & haha & rock&us & different & women & tumblr\\
running & heart & taking & curious&ambitious & share & research & spontaneous\\
\hline
\end{tabular}
\end{center}
\caption{\label{font-table} The top ten words for each of the MBTI traits. }
\end{table*}

The word lists indicate that topic modeling was successful in finding words that describe the different personality traits. Though I couldn't find any previous research which developed word lists such as these there is plenty of work which has focused on the relationship between linguistic cues and personality types. Some of these cues include the number of self references, emotion words, negations, inclusive/exclusive words, different parts of speech, and sentence length \citep{mairesse2007using,Argamon05lexicalpredictors}. These cues have been found to correlate with some of the big five personality traits \citep{pennebaker1999linguistic}.

As an evaluation our MBTI trait specific word lists are compared with the known linguistic cues of personality. The hypothesis is that if these word lists indeed correlate with the different personality traits then certain lists should also correlate with the known linguistic indicators. Using the non-labeled random dataset several features were extracted and then correlated with one another. Features describing the previously stated linguistic cues were measured using their counts and normalized by the total number of words in the profile. Four features using the words lists were computed using the ratio of the counts between the traits of each dichotomy. For example, the feature representing attitude was calculated using the ratio between words found on the Extraversion list and words found on the Introversion list. Similarly, a ratio for the counts of positive and negative emotion words was also computed.

Each feature was correlated with one another and a false discovery rate control was used to correct for the multiple comparisons. The results are shown in Table 2.

\begin{table}[h]
\begin{center}
\begin{tabular}{|l|l|l|}
\hline \bf Feature 1 &\bf Feature 2 &\bf $R\textsuperscript{2}$  \\ \hline
Attitude (E/I) & Judging (T/F) & -.10* \\
Judging & Avg. Sent Len. & -.15*** \\
Attitude & Exlusive words  & {\color[rgb]{0,0,1}-.11**}\\
Attitude & Avg. Sent Len. & {\color[rgb]{0,0,1}.10*} \\
Emotion Words & Avg. Sent Len. & {\color[rgb]{0,0,1}.43***} \\
Self References & Emotion Words & {\color[rgb]{0,0,1}.11*} \\
Pos/Neg Emotion & Avg. Sent Len. & {\color[rgb]{1,0,0}.11*}\\
Self References & Avg. Sent Len. & {\color[rgb]{0,0,1}.54***}\\
\hline
\end{tabular}
\end{center}
\caption{Correlations between different features. Blue indicates a correlation which supports the literature and red indicates a contradiction.}
\end{table}

There were five significant correlations between the MBTI word list features and linguistic indicators. The attitude and judging features correlated with sentence length. This means introverts tend to write longer sentences, which is supported by the literature \citep{mehl2006personality} and those with a Thinking-type judging function also write longer sentences. The other significant correlation was between attitude and exclusive words (but, without, exlude, he, she). People that score high on extraversion tend to use fewer exclusive words. This is also supported by the literature \citep{pennebaker1999linguistic}.

The one contradiction came between average sentence length and the positive/negative emotion words ratio. Introverts tend to use negative words at a higher rate than extroverts so a positive correlation with sentence length should be expected. We found a negative correlation, however, though the correlation is significant it is not very strong.

\section{Personality Specific Preferences}
\subsection{Methods}
The second objective is to find word lists that describe aesthetic preferences of each MBTI trait. A similar procedure as the previous section was carried out. The MBTI-labeled dataset was split according to the traits and trait specific topic modeling was performed. However, this time the content consisted of the ``favorite books, movies, shows, music, and food” portion of each profile. The generated word lists consist of words describing television show titles, band names, genres, food, and other favorites. The complete lists for each trait are shown in Figures 5-8 of the appendix.

To evaluate the preference word lists the non-labeled dataset of 516 random profile was used. The hypothesis is that features extracted from the ``favorite books, movies, shows, music, and food” section using these trait specific preference word lists should correlate with their counterpart trait specific features extracted from the rest of the user's profile. For example, a profile that exhibits a higher level of extraversion in the personality sections should exhibit an increase of preferences which correspond to the extrovert word list generated in this section.

To test this hypothesis we calculated eight features on the non-preference sections of the profiles. The extraversion feature was calculated by computing the average number of words from the extrovert word list per sentence. The introversion feature was calculated by computing the average number of words from the introvert word list per sentence and so on for the remaining six MBTI traits. Similarly, eight preference features on the "favorites" section were calculated using the eight preference word lists. Correlations were then measured between different pairs of personality features and preference features. The p-values were again corrected using a false discovery rate control.

\subsection{Results}
The correlations from the experiment are shown below in Table 3.
\begin{table}[h]
\begin{center}
\begin{tabular}{|c|l|}
\hline \bf Personality-Preference &\bf  $R\textsuperscript{2}$  \\ \hline
E - E & .15*** \\
I - I & .13** \\
N - N & .15***\\
T - T & .14** \\
J - J & .18***\\
P - P& .11**\\
\hline
\end{tabular}
\end{center}
\caption{Correlations between different personality and preference features. }
\end{table}

Our hypothesis was true for six of the eight pairs. The two pairs consisting of the sensing and feeling traits did not yield a significant correlation.

\section{Preference Topics and Personality Traits}
\subsection{Method}
The final goal is to find topics that occur in the preferences section over all the personality types and test whether any of the topics correlate with particular traits. To do this LDA of the preferences was performed over all the profiles. A total of 20 topics were created. Therefore, 20 preference features were extracted from the profiles using the generated topic word lists and eight personality features were extracted using the MBTI trait word lists (using the same method as described in the previous section). The preference topics and personality features were then correlated and multiple tests were controlled using the false discovery rate method.

\subsection{Results and Discussion}
There were signicant correlations for each of the twenty topics. Five topics are presented below. These topics were identified as higher-level categories. These hand-labeled topics, the top topic words, and traits they correlate with are shown in Table 4.

\begin{table}[h]
\begin{center}
\begin{tabular}{|p{1.8cm}|p{3cm}|p{2cm}|}
\hline \bf Topic &\bf  Top Words &\bf Correlated Traits\\ \hline
Sci-Fi/Fantasy & star, rings, lord, doctor, firefly,who, series, potter, trek, rock & N, T \\
General stuff loved by all & fiction, man, ice, michael, castle, cream, beer, dance, cake, jackson, & E-I, S-N, T-F, J-P \\
Popular TV Shows & breaking, bad, thrones, game, arrested, development, sunny, 30, parks, lost & E-I, N, T, J\\
Intl Cuisine & thai, indian, mexican, japanese, korean, sushi, folk, fiction, metal, indie & N, T-F, P\\
Upbeat Dance Music & rap, pop, music, dance, hip, band, hop, especially, listen, house & E, S-N, F, J-P\\
\hline
\end{tabular}
\end{center}
\caption{Significant positive correlations between preference topics and personality traits.}
\end{table}

Some of the correlations found here are reaffirmed by previous studies. Rentfrow and Gosling found that extraversion and aggreeableness correlated positively with energetic, rhytmic music \citep{rentfrow2003re}. The Extraversion and Feeling MBTI traits correspond to these two big five traits respectively and were found to correlate positively with the ``upbeat dance music" topic. 

Considering that all of these profiles came from within the United States it is reasonable to think that the big five trait of oppenness to experience might correlate positively with international cuisine. This is indeed the case. The iNtuition MBTI trait corresponds to being more open and correlates positively with the international cuisine topic. Rentrfrow and Gosling also found that reflective and complex music (such as jazz and folk) as well as intense and rebellious music (metal, alternative) correlated positively with oppenness to experience \citep{rentfrow2003re}. The music genres of folk, metal, and indie were also present in the cuisine topic model supporting their findings.

Overall topic modeling and correlation measurements seemed successful. Several signficant correlations were found and many of the findings were supported by previous research in the field. However, the correlations that were found were not particularly strong and there was quite a bit of preference overlap among the different personality traits. Additionally, no significant negative correlations were found. This suggests that many preferences are shared over the personalities and there are slight tendencies for people with specific traits to also have preferences for certain music, shows, or food types.

\section{Conclusion and Future Work}
In this work we were able to identify words from Okcupid profiles that are good indicators of MBTI personality traits. We also identified preferences that correlate with these traits. Finally, we found topics of preferences and found significant correlations between the topics and certain traits. A natural next step for this work would be take a more engineering approach and train classifiers to predict the different personality and preference types. Additionally, in predicting preferences it might not be the case that the personality features are the most optimal. It would be interesting to look for non-personality related topics that correlate with preferences. For example, during data collection we saw that individuals across several different personality types stated they played instruments. The fact that one plays an instrument might correlate with the type of music they listen to.

It is also worth noting some other strategies we attempted in our analysis. We used NLTK to find bigram and trigram collocations of the preferences. Finding reasonable collocations such as "breaking bad", "game of thrones", and "michael jackson" worked well, but the correlation results were unaffected. However, such collocations can be useful in other regards. This dataset is not extremely large so reducing the feature size can be helpful and using the collocations to do this may be possible. For example, many profiles mentioned specific bands such as "arcade fire" or "imagine dragons" and others simply mentioned genres such as "indie" or "alternative."

Additionally, we tried MBTI label specific topic modeling (ENTJ, ISTP, etc.) of the preferences. This method was motivated by assumption that different preferences unique to each of the 16 labels would appear, however this was not the case. Instead of the preferences such as band names and tv shows appearing in each MBTI label, filler words unique to each label appeared. The words didn't describe the preferences, but instead seemed to describe how each personality type describe their preferences. For example, one type might tend to write ``I really like ..." and another type might tend to write ``I fucking love..." It would probably be beneficial to add these words to the personality trait word lists.

\bibliographystyle{named}
\bibliography{main}



\section{Appendix}
\newcommand*{\figuretitle}[1]{%
    {\centering
    \textbf{#1}
    \par\medskip}
}

\begin{figure*}[h]
\figuretitle{Extraversion vs Introversion}
  \includegraphics[width=\textwidth]{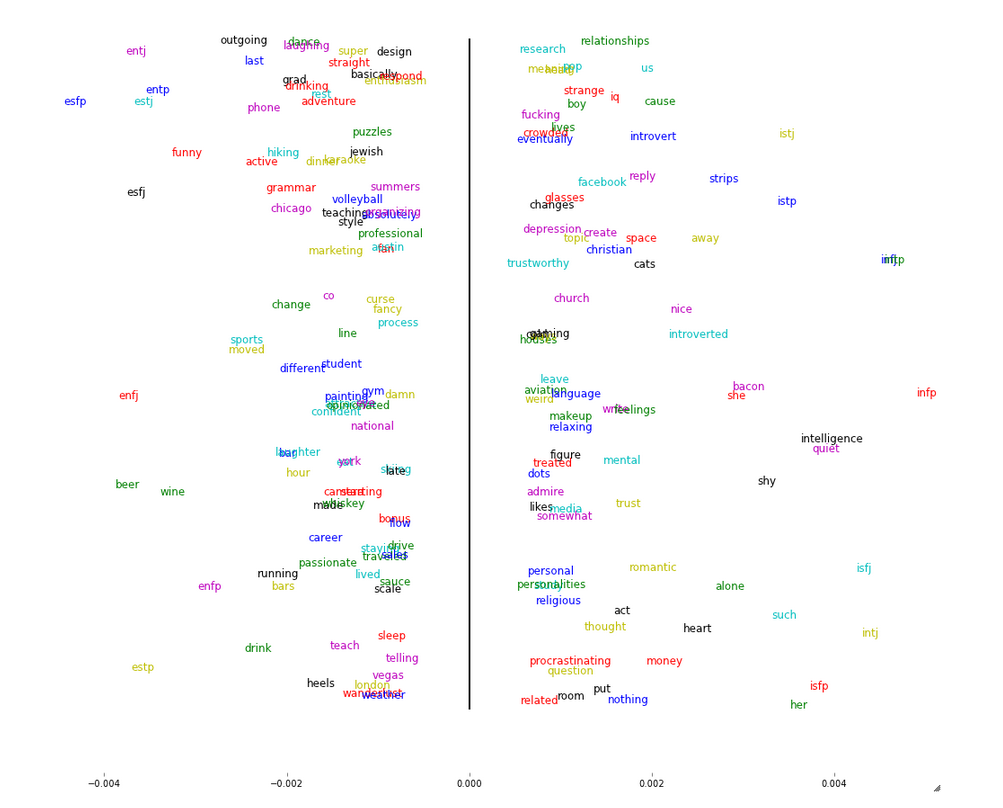}
  \caption{The top profile words for extraverison and introversion generated by personality specific LDA. }
\end{figure*}

\begin{figure*}[h]
\figuretitle{Sensing vs Intuition}
  \includegraphics[width=\textwidth]{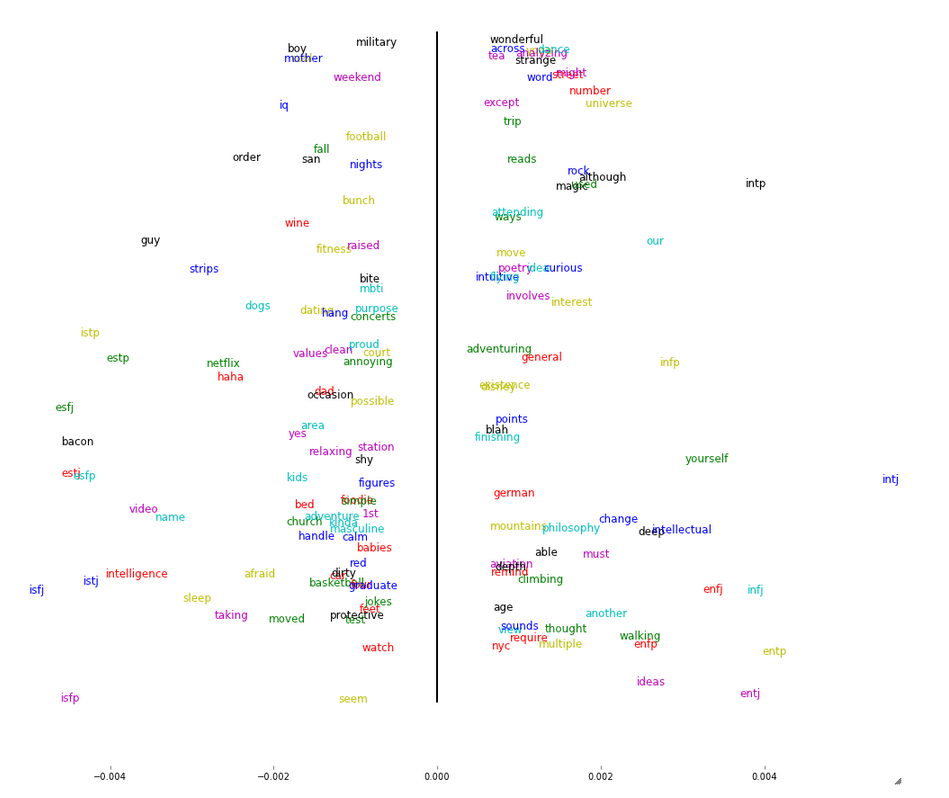}
  \caption{The top profile words for Sensing and Intuition generated by personality specific LDA. The big five trait of "oppenness" is shown to positive correlated with Intuition and negatively with Sensing}
\end{figure*}

\begin{figure*}[h]
\figuretitle{Thinking vs Feeling}
  \includegraphics[width=\textwidth]{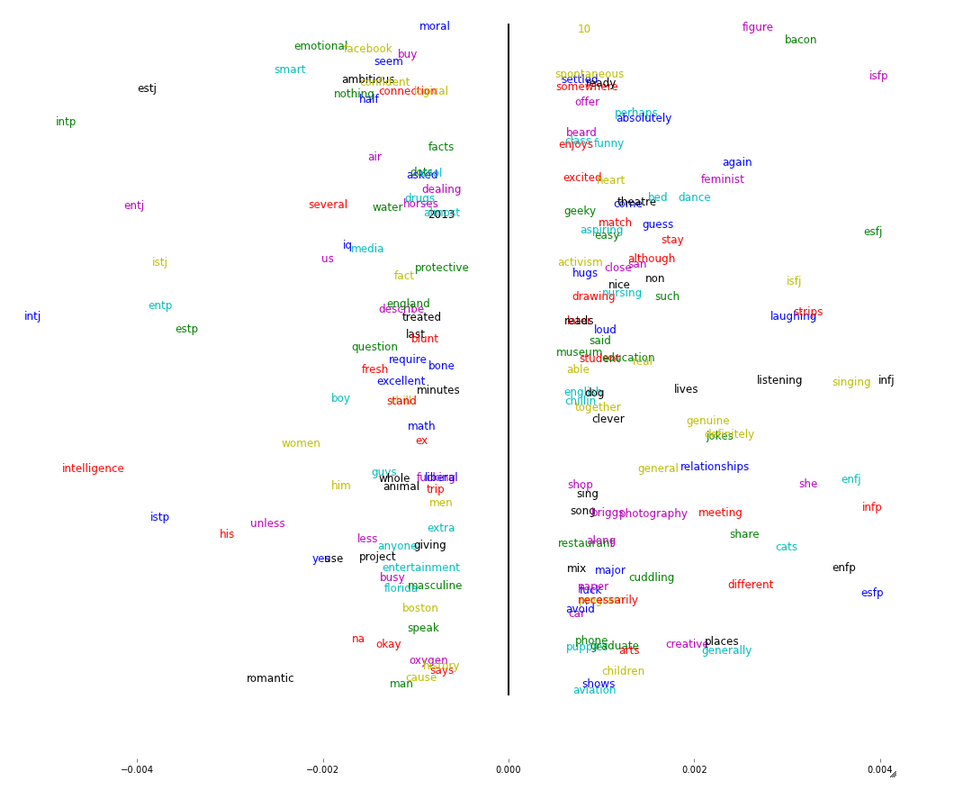}
  \caption{The top profile words for Thinking and Feeling generated by personality specific LDA. The big five trait of "agreeableness" is shown to positive correlated with Feeling and negatively with Thinking}
\end{figure*}

\begin{figure*}[h]
\figuretitle{Judging vs Perception}
  \includegraphics[width=\textwidth]{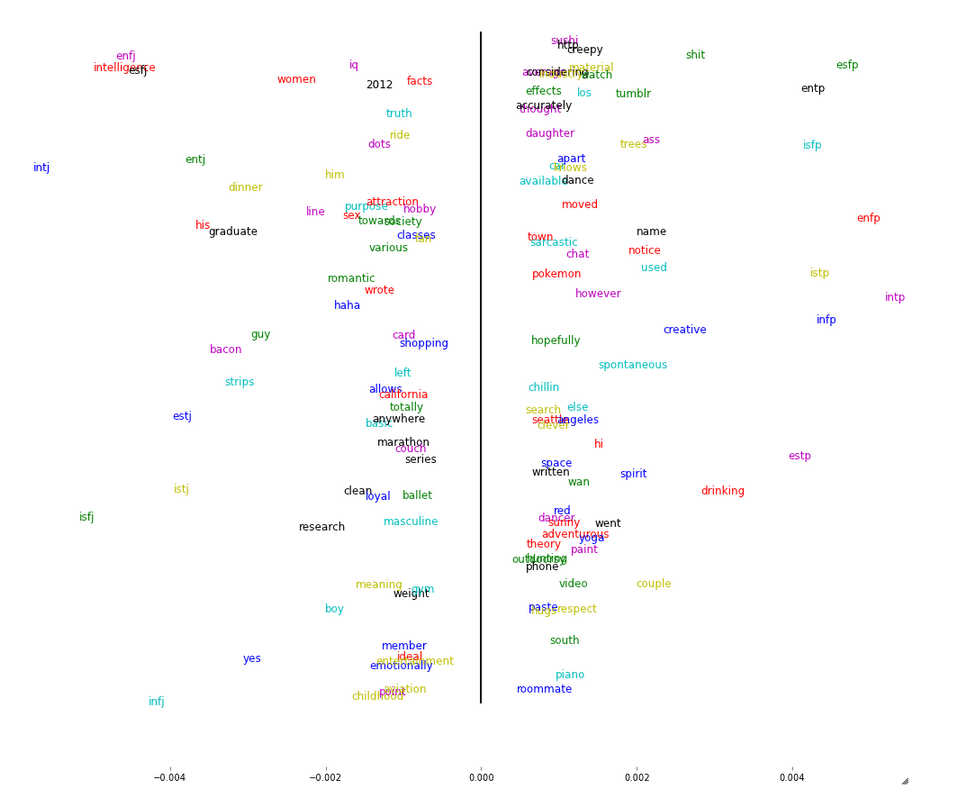}
  \caption{The top profile words for Judging and Perception generated by personality specific LDA. The big five trait of "conscientiousness" is shown to positive correlated with Judging and negatively with Perception}
\end{figure*}

\begin{figure*}[h]
\figuretitle{Extraversion vs Introversion}
  \includegraphics[width=\textwidth]{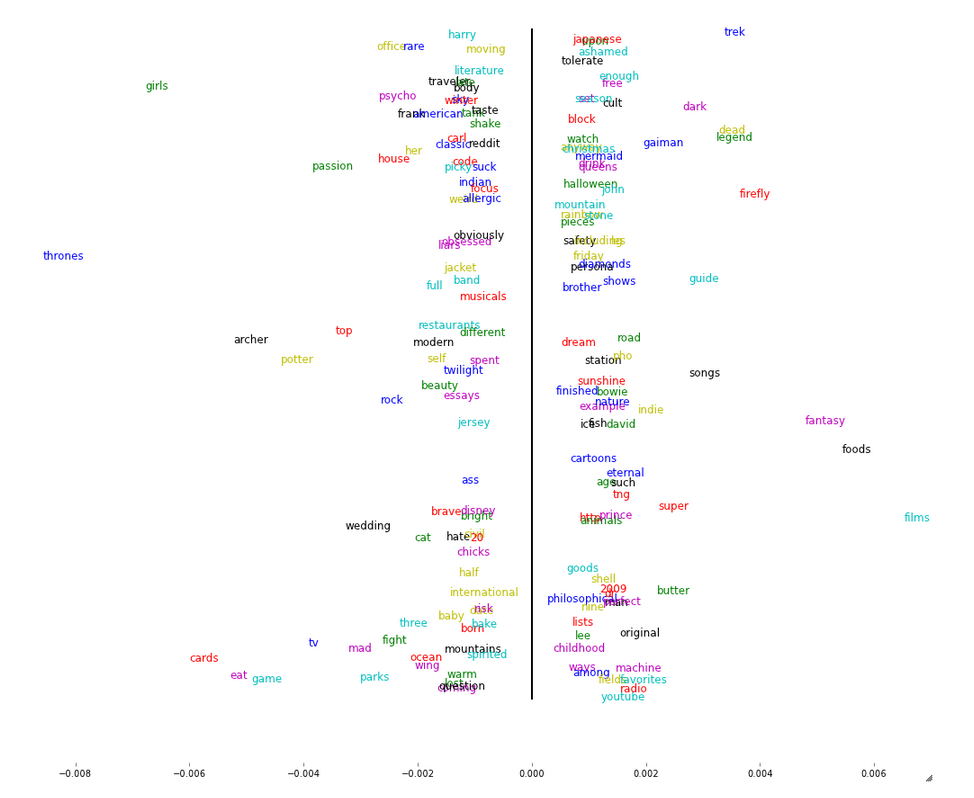}
  \caption{The top preference words for extraverison and introversion generated by personality specific LDA. }
\end{figure*}

\begin{figure*}[h]
\figuretitle{Sensing vs Intuition}
  \includegraphics[width=\textwidth]{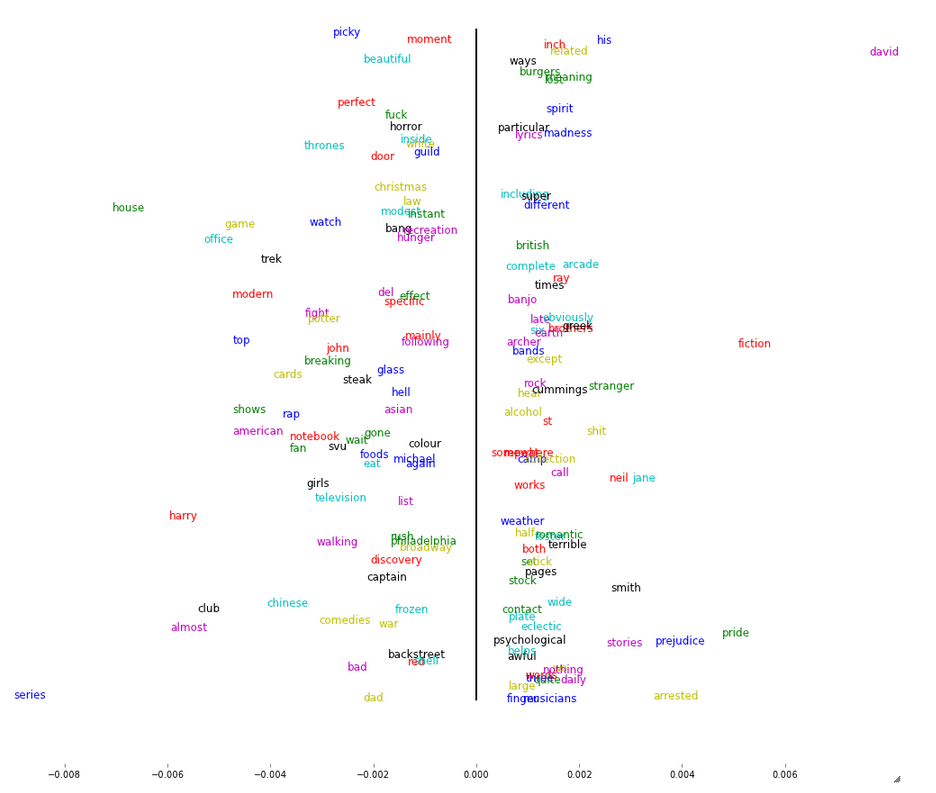}
  \caption{The top preference words for Sensing and Intuition generated by personality specific LDA. The big five trait of "oppenness" is shown to positive correlated with Intuition and negatively with Sensing}
\end{figure*}

\begin{figure*}[h]
\figuretitle{Thinking vs Feeling}
  \includegraphics[width=\textwidth]{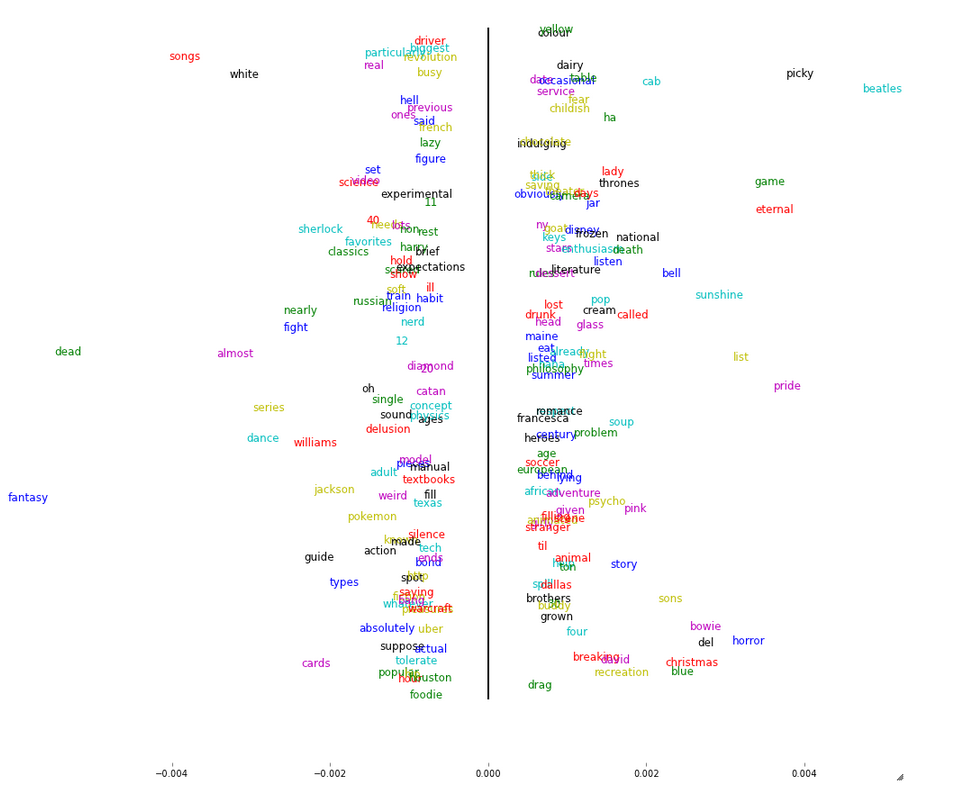}
  \caption{The top preference words for Thinking and Feeling generated by personality specific LDA. The big five trait of "agreeableness" is shown to positive correlated with Feeling and negatively with Thinking}
\end{figure*}

\end{document}